\input jytex.tex   % available from hep-th 
%\draft 
\typesize=10pt \magnification=1200 
\baselineskip17truept %\baselineskip25truept 
\hsize=6truein\vsize=8.5truein %\leftmargin=1.25in 
%\oddleftmargin=.5in 
%\evenleftmargin=1.5in
\sectionnumstyle{blank}
\chapternumstyle{blank}
\chapternum=1
\sectionnum=1
\pagenum=0
%\input squash.lab
%\referencestyle{preordered}
% title style follows

\def\begintitle{\pagenumstyle{blank}\parindent=0pt\begin{narrow}[0.4in]}
\def\endtitle{\end{narrow}\newpage\pagenumstyle{arabic}}

% exercise style follows

\def\beginexercise{\vskip 20truept\parindent=0pt\begin{narrow}[10 
truept]}
\def\endexercise{\vskip 10truept\end{narrow}}

% **************    my jyTeX abbreviations   *****************

\def\eql#1{\eqno\eqnlabel{#1}}
\def\ref{\reference}
\def\peq{\puteqn}
\def\pref{\putref}

\def\mgn{\marginnote}
\def\bex{\begin{exercise}}
\def\eex{\end{exercise}}

% *********************** My definitions ************************  

 %scaled\magstep1 % For VAX. Borde p195.
 %scaled\magstep1 % For VAX. Borde p195.
%\font\open=msym10 %scaled\magstep1 % For Arbortxt on PC
%\font\opens=msym8 %scaled\magstep1 % For Arbortxt on PC
%\font\goth=eufm10  % For Arbortxt on PC, and VAX. Borde p199
%\font\ssb=cmss10 
%\font\smsb=cmss8 
\def\mbox#1{{\leavevmode\hbox{#1}}}

\def\hspace#1{{\phantom{\mbox#1}}}

\def\al{\alpha}
 %in jyTeX
 %in jyTeX
 %in jyTeX
 %in jyTeX
 %in jyTeX
 %in jyTeX
% in jyTeX
% in jyTeX
% in jyTeX
\def\be{\beta}
\def\ga{\gamma}
\def\de{\delta}
\def\Ga{\Gamma}

\def\la{\lambda}
\def\La{\Lambda}
\def\om{\omega}

\def\si{\sigma}

\def\th{\theta}

\def\ze{\zeta}

\def\De{\Delta}

\def\Real{{\rm Re\,}}

\def\Imag{{\rm Im\,}}

\def\zf{$\zeta$--function}
\def\zfs{$\zeta$--functions}

     % Newline

\def\frac#1/#2{\leavevmode\kern.1em
\raise.5ex\hbox{\the\scriptfont0 #1}\kern-.1em/\kern-.15em
\lower.25ex\hbox{\the\scriptfont0 #2}}
\def\sfrac#1/#2{\leavevmode\kern.1em
\raise.5ex\hbox{\the\scriptscriptfont0 #1}\kern-.1em/\kern-.15em
\lower.25ex\hbox{\the\scriptscriptfont0 #2}}

\def\gtorder{\mathrel{\raise.3ex\hbox{$>$}\mkern-14mu
             \lower0.6ex\hbox{$\sim$}}}
\def\ltorder{\mathrel{\raise.3ex\hbox{$<$}\mkern-14mu
             \lower0.6ex\hbox{$\sim$}}}

\def\semidirprod{\rlap{\ss C}\raise1pt\hbox{$\mkern.75mu\times$}}
\def\for{\lower6pt\hbox{$\Big|$}}
\def\fish{\kern-.25em{\phantom{abcde}\over \phantom{abcde}}\kern-.25em}

 %triple 
%dot
 %double 
%dot
 %double dot 
%for small #1

\def\boxit#1{\vbox{\hrule\hbox{\vrule\kern3pt
        \vbox{\kern3pt#1\kern3pt}\kern3pt\vrule}\hrule}}
\def\dalemb#1#2{{\vbox{\hrule height .#2pt
        \hbox{\vrule width.#2pt height#1pt \kern#1pt
                \vrule width.#2pt}
        \hrule height.#2pt}}}

        %double stroke
\def\frac#1#2{{{#1}\over{#2}}}
 %lower covariant deriv. 
    %lower ordinary  deriv.

      %Connection
    %Connection'

\def\etc{{\it etc. }}

\def\eg{{\it e.g. }}
\def\ie{{\it i.e. }}

 %gives average <#1>
 %gives thermal average <<#1>>
   %gives bracket <#1|#2>
 %gives big bracket <#1|#2>
  %gives 
%matrix element <#1|#2|#3>

\def\Tr{{\rm Tr\,}}

\def\sumdasht#1#2{{\mathop{{\sum}'}_{#1}^{#2}}}

\def\3j#1#2#3#4#5#6{\left\lgroup\matrix{#1&#2&#3\cr#4&#5&#6\cr}
\right\rgroup}

\def\cac{{\cal C}}

\def\man{{\cal M}}

\def\m?{\mgn{?}}
% KK's defs

\def\beq{\begin{eqnarray}}
\def\eeq{\end{eqnarray}}

%  *******************  Journal refs **********************

\def\aop#1#2#3{{\it Ann. Phys.} {\bf {#1}} (19{#2}) #3}

\def\cmp#1#2#3{{\it Comm. Math. Phys.} {\bf {#1}} (19{#2}) #3}
\def\cqg#1#2#3{{\it Class. Quant. Grav.} {\bf {#1}} (19{#2}) #3}

\def\jgp#1#2#3{{\it J. Geom. and Phys.} {\bf {#1}} (19{#2}) #3}
\def\jmp#1#2#3{{\it J. Math. Phys.} {\bf {#1}} (19{#2}) #3}
\def\jpa#1#2#3{{\it J. Phys.} {\bf A{#1}} (19{#2}) #3}

\def\np#1#2#3{{\it Nucl. Phys.} {\bf B{#1}} (19{#2}) #3}
\def\pl#1#2#3{{\it Phys. Lett.} {\bf {#1}} (19{#2}) #3}

\def\prD#1#2#3{{\it Phys. Rev.} {\bf D{#1}} (19{#2}) #3}

\def\cras#1#2#3{{\it Comptes Rend. Acad. Sci. (Paris)} {\bf{#1}} (#2) #3}

\def\mpcps#1#2#3{{\it Math. Proc. Camb. Phil. Soc.} {\bf{#1}} (19{#2}) #3}

\def\am#1#2#3{{\it Acta Mathematica} {\bf {#1}} (19{#2}) #3}
\def\aim#1#2#3{{\it Adv. in Math.} {\bf {#1}} (19{#2}) #3}
\def\ajm#1#2#3{{\it Am. J. Math.} {\bf {#1}} ({#2}) #3}

\def\aom#1#2#3{{\it Ann. of Math.} {\bf {#1}} (19{#2}) #3}

\def\cpde#1#2#3{{\it Comm. Partial Diff. Equns.} {\bf {#1}} (19{#2}) #3}

\def\invm#1#2#3{{\it Invent. Math.} {\bf {#1}} (19{#2}) #3}
\def\ijpam#1#2#3{{\it Ind. J. Pure and Appl. Math.} {\bf {#1}} (19{#2}) #3}
\def\jdg#1#2#3{{\it J. Diff. Geom.} {\bf {#1}} (19{#2}) #3}

\def\jmpa#1#2#3{{\it J. Math. Pures. Appl.} {\bf {#1}} ({#2}) #3}

\def\ojm#1#2#3{{\it Osaka J.Math.} {\bf {#1}} ({#2}) #3}

\def\pja#1#2#3{{\it Proc. Jap. Acad.} {\bf {A#1}} (19{#2}) #3}

\def\tams#1#2#3{{\it Trans. Am. Math. Soc.} {\bf {#1}} (19{#2}) #3}

% *******************   Main text *********************
%\begin{ignore}
\begin{title}  
\vglue 1truein
%\righttext {ICTP/23}
%\righttext{hep-th/98}
\vskip15truept
%\leftline{\today}
%\vskip 30truept
\centertext {\Bigfonts \bf Effective actions on the squashed three-sphere}
\vskip 20truept 
\centertext{J.S.Dowker\footnote{{Department of Theoretical Physics,
The University of Manchester, Manchester, England}}}
\centertext{{\it Theory Group, Department of Physics, Imperial College,}}
\centertext{{\it Blackett Laboratory, Prince Consort Rd, London.}}
\vskip 20truept
\centertext {Abstract}
\vskip10truept
\begin{narrow}
The effective actions of a scalar and massless spin-half field are
determined as functions of the deformation of a symmetrically 
squashed three-sphere. The extreme oblate case is particularly
examined as pertinant to a high temperature statistical mechanical
interpretation that may be relevant for the holographic principle.
Interpreting the squashing parameter as a temperature, we find that
the effective `free energies' on the three-sphere are mixtures of 
thermal two-sphere scalars and spinors which, in the case of the spinor 
on the three-sphere, have the `wrong' thermal periodicities.
However the free energies do have the same leading
high temperature forms as the standard free energies on the two-sphere.
The next few terms in the high-temperature series are also evaluated
and briefly compared with the Taub-Bolt-AdS bulk result.
\end{narrow}
\vskip 5truept
\righttext {December 1998}
\vskip 60truept
%\righttext{Typeset in \jyTeX}
\vfil
\end{title}
\pagenum=0
%\end{ignore}
\section{\bf 1. Introduction}

The holographic `principle' says, in its barest form, that the
information contained in the interior of a space-time domain is 
encoded in a field theory residing on the boundary. In accordance
with this, the gravitational entropies of the Taub-Nut-AdS and
Taub-Bolt-AdS space-times have been computed by Hawking, Hunter
and Page [\pref{HHP}] and by Chamblin {\it et al} [\pref{CEJM}] 
with the aim of comparing with boundary (conformal) field theories. 
The boundaries are symmetrically squashed three-spheres, possibly 
with identifications.

The object of this paper is to detail the calculation of the
effective action, \ie the functional determinants, of scalar and
spinor fields as functions of the squashing on the three-sphere. The 
effective action will be associated with a free energy and thence with 
an entropy. 

The fields will be free so this is only a prelude to a more realistic
investigation and the present results may well have only a passing
relevance to the holographic principle. Such a stopgap calculation 
has been suggested by Hawking {\it et al} in [\pref{HHP}]. 
Nevertheless, the evaluation of the determinants is of interest in 
itself.  Our results here will be confined to the determinants. 
Elaboration of the statistical mechanical interpretation is left for 
a later paper although some preliminary comparison with the 
bulk Taub-Bolt-AdS result is made.

\section{\bf 2. The basic situation}

The squashed 3-sphere appears as the spatial section of the frozen 
mixmaster universe and quantum field theory on this space-time was
discussed by Hu, Fulling and Parker [\pref{HFP}], Hu [\pref{Hu}], 
Shen, Hu and O'Connor [\pref{SHC}] and 
Critchley and Dowker [\pref{CrandD}]. It has 
been discussed in a Kaluza-Klein setting by Okada [\pref{Okada}] and by 
Shen and Sobczyk [\pref{SandS}].

A discussion of the vacuum energy of a massless spin-half field on the 
squashed 3-sphere has been given in [\pref{dow11}] and for this reason
we will reexamine this case before looking at the scalar field.

In  reference [\pref{dow11}], in order that rules of  standard 
angular momentum 
theory should apply unmodified it was necessary to choose the
radius of the unsquashed sphere, $S^3$, to be $a=2$. 
and the (standard) metric in Euler angles is
$$
ds^2=(d\th^2+\sin^2\th d\phi^2)+l_3^2(d\psi+\cos\th d\phi)^2
\eql{3metric}$$
showing the (symmetrically) squashed $S^3$ as a twisted product, 
$S^2\times S^1$. The circle has radius $2l_3$, and hence
a circumference of $4\pi l_3$. If, illustratively, this periodicity is 
translated into a temperature, we find $\be=1/T=4\pi l_3$. Another way
of saying this is to note that the range of $\psi$ is $0\to4\pi$ and 
then to interpret $l_3\psi$ as a Euclidean time. We will be 
particularly interested in the extreme oblate case, $l_3\to0$, when
the metric reduces to that on the unit two-sphere. The relation between
the \zfs\ will constitute a type of twisted Selberg-Chowla formula. 

In the notation of Hawking, Hunter and Page [\pref{HHP}], $l_3^2=E$. 
It is possible to identify points on the `$\psi$-circle', and this is 
the more interesting situation. Nevertheless it will not be 
considered here. The work [\pref{CEJM}] also treats only this
simplest case. In the notation of [\pref{CEJM}], $l_3^2=4n^2/l^2$. 
We denote the squashed sphere by $\widetilde S^3$.

\section{\bf 3. Spinor \zfs\ on $\widetilde S^3$.}

We ignore all restrictions on the boundary spinor theory arising from 
its embedding in some bulk theory.
For odd-dimensional spaces the relevant operator is thus the Pauli one
which, on the squashed three-sphere, is $\Pi=-i\si^i\nabla_i$
in terms of (covariant) Pauli matrices and the spinor covariant
derivatives, which we will not exhibit here. The dimension of spinor
space on $\widetilde S^3$ is 2, which is the same as that on $S^2$.

The eigenvalues of $\Pi$ are determined to be (Hitchin [\pref{Hitch}], 
Gibbons [\pref{Gibb}], and [\pref{dow11}])
$$
\om_{\pm}=(2l_3)^{-1}
\bigg({1\over2}l_3^2\pm\big(n^2+4(l_3^2-1)q(n-q)\big)^{1/2}\bigg)
\eql{eigenv}$$
where the integers $n$ and $q$ emerge from the angular momentum quantum
numbers labelling the unperturbed states (before the necessary 
secular diagonalisation). Thus $n=2L+1$ and $q=n/2-M$ with $L$ an
orbital label and $M$ the projection of the total angular momentum.
 
If $l_3<4$, $\om_+$ is positive and $\om_-$ negative, and for ease 
we will assume that this is so. The corresponding traced \zfs\ are 
constructed separately,
$$
\ze_+(s)=\sum_{n=1}^\infty\sum_{q=0}^n{n\over\om_+^s}
$$
and
$$
\ze_-(s)=\sum_{n=2}^\infty\sum_{q=1}^{n-1}{n\over(-\om_-)^s}
$$
which exhibit the quantum number range restrictions with $n$ the
remaining degeneracy of the modes. We note that, as a function of $l_3^2$,
nothing peculiar happens to the eigenvalues as $l_3$ passes through 1 
and also that there is a square root branch point at $l_3^2=0$.

The \zf\ for the squared operator $\Pi^2$ is 
$$
\Tr_3\ze_3(s)=\ze_+(2s)+\ze_-(2s).
$$

The awkwardness of the eigenvalues, (\peq{eigenv}), restricted the
analysis in [\pref{dow11}] to a power series expansion in the 
squashing parameter $l_3$.
For the time being we shall continue with this expansion which,
in any case, is adequate for the high temperature limit.
The computation then reduces to that of the function, $f(s)$, defined, 
for $\Real s>3/2$, by
$$
f(s)=\sum_{n=2}^\infty\,\sum_{q=1}^{n-1} {n\over{\big(n^2
+4\de^2q(n-q)\big)^s}},\quad{\rm where}\,\,\de^2=l_3^2-1.
\eql{sumdef}$$

In terms of $f(s)$, $\Tr_3\ze_3(s)$ reads,
$$
\Tr_3\ze_3(s)=2(2l_3)^s\bigg(\ze(2s-1,l_3^2)-w\ze(2s,l_3^2)+f(s) + 
l_3^4s(2s+1)
f(s+1)+ O(l_3^8)\bigg).
\eql{expn}
$$

This expansion is valid, numerically for $l_3<\sqrt2$. One can rejig the
series to allow for larger values, but we will not bother. With this
approach, it is not possible to discuss the $l_3\to\infty$ limit.

A preliminary aim is to calculate the (Euclidean) effective action, 
which we define to be $\Tr_3\ze_3'(0)/2\equiv W_{\rm sp}$. 

Noting that there is no conformal anomaly, $\Tr_3\ze_3(0)=0$,
the complete formal series is
$$
W_{\rm sp}=2\ze'(-1,l_3^2)-2l_3^2\ze'(0,l_3^2)+f'(0)+
l_3^4(2P+R)+
{1\over4}l_3^8f(2)+O(l_3^{12})
\eql{detexpsn}$$
where, since $f(s)$ has a pole at $s=1$, we have to write
$$
f(s)={P\over s-1}+R+O(s-1).
$$

It is expected that
the interesting behaviour as $l_3$ becomes small is contained 
in $f'(0)$ and we must determine the analytic continuation
of this quantity. Initially we are looking for terms which diverge
as $l_3\to 0$.

In [\pref{dow11}] we employed the standard Plana summation formula and,
for convenience, will proceed in the same way although there are other
approaches.

Extending again the $q$ sum to $0$ to $n$ in (\peq{sumdef}), an 
application of the Plana summation formula yields, for deformations
in the prolate direction, [\pref{dow11}],

$$\eqalign{
f(s)=&-\ze_R(2s-1)+\ze_R(2s-2)\int_0^1{dy\over\big(1+4\de^2y(1-y)\big)^s}
\cr
&-2i\int_0^\infty {dt\over \exp(2\pi t)-1}\bigg\{\sum_{n=1}^\infty
{n\over\big(n^2+4\de^2(t^2-itn)\big)^s}-(t\to -t)\bigg\}\cr}
\eql{Planaq}
$$
which is not a complete continuation as the $n$ summation has yet to be
dealt with. The pole at $s=1$ is apparent giving $P=-1/2$.

For deformations in the oblate direction ($0\le l_3\le1$), which we need
for the high temperature series, singularities of the summand
encroach into the relevant part of the complex $q$ plane, \ie
$0\le\Real q\le n$. Concentrating now on this oblate case, `extra'
branch points occur at 
$$
q={n\over2}\pm i{n\over2}{l_3\over\sqrt{1-l_3^2}}
\equiv{n\over2}\big(1\pm i\tan\th\big)
$$
and we give the details of the resulting contributions. Following the 
usual procedure (as in Lindel\"of [\pref{Lindel}]) we find (setting $b^2=-\de^2$ and 
omitting the $n$-summation for the moment) the additional pieces,
$$
\int_U {dz\over \exp(-2\pi i z)-1}{n\over\big(n^2-4b^2z(n-z)\big)^s}+
\int_L {dz\over \exp(2\pi i z)-1}{n\over\big(n^2-4b^2z(n-z)\big)^s}
$$
where the infinite $U$ contour runs {\it anti}clockwise around the cut 
from $n\,(1+ i \tan\th)/2$ to ${n/2}+i\infty$, while $L$
runs clockwise around the corresponding cut in the lower half plane.
Symmetry means that the two contributions are equal and we get, after 
a change of variable, $z=(1+i\ze\tan\th)\,n/2$, the extra contribution
to $f(s)$,
$$
{i\tan\th\over l_3^{2s}}\sum_{n=1}^\infty{1\over n^{2s-2}}
\int_{C} {d\ze\over \big((-1)^n\exp(n\pi\ze\tan\th)-1\big)(1-\ze^2)^s}
\eql{xtraint}$$
where $C$ is an anticlockwise contour running around 
the real $\ze$-axis, cut from 1 to $\infty$.
For $s$ a non-negative integer, this expression vanishes, while,
if $s$ is a negative integer, it can be evaluated using residues,
providing a useful numerical check.

Because of the exponential factor in the denominator, expression
(\peq{xtraint}) converges for all $s$. It can be taken as the 
analytic
continuation and should be added to (\peq{Planaq}), for deformations
in the oblate direction, to give
$$\eqalign{
f(s)=&-\ze_R(2s-1)+\ze_R(2s-2)\int_0^1{dy\over
\big(1-4b^2y(1-y)\big)^s}\,-\cr
&2i\int_0^\infty {dt\over \exp(2\pi t)-1}\bigg\{\sum_{n=1}^\infty
{n\over\big(n^2-4b^2(t^2-itn)\big)^s}-(t\to -t)\bigg\}\,+\cr
&{i\tan\th\over l_3^{2s}}\sum_{n=1}^\infty{1\over n^{2s-2}}
\int_{C} {d\ze\over \big((-1)^n
\exp(n\pi\ze\tan\th)-1\big)}{1\over\big(1-\ze^2\big)^s},\cr}
\eql{Planaqo}
$$

We shall denote the last two terms on the right-hand side of 
(\peq{Planaqo}) by $f_2(s)$ and $f_3(s)$ respectively and consider
$f_3$ first.

If $s$ is such that the integral converges at $\ze=1$, 
(\peq{xtraint}) can be 
converted to a real integral in the usual way by combining the upper
and lower pieces of $C$. One gets, choosing appropriate phases,  
$$
f_3(s)=2\sin(\pi s){\tan\th\over l_3^{2s}}\sum_{n=1}^\infty{1\over 
n^{2s-2}}\int_1^\infty {d\ze\over \big((-1)^n
\exp(n\pi\ze\tan\th)-1\big)}{1\over\big(\ze^2-1\big)^s},
\eql{xtraint2}$$

In particular we can differentiate with respect to $s$ to get the 
contribution to $f'(0)$,
$$
f_3'(0)=2\pi\tan\th\sum_{n=1}^\infty n^2\int_1^\infty 
{d\ze\over \big((-1)^n\exp(n\pi\ze\tan\th)-1\big)}.
\eql{diff0}$$

This is a simple example of a Lerch function. Expansion of the
integrand allows the integration to be done to yield, from 
(\peq{detexpsn}),  the contribution to ${1\over2}\Tr_3\ze_3'(0)$,
$$
W^{(3)}_{\rm sp}=2\sum_{m=1}^\infty {1\over m} 
\sum_{n=1}^\infty (-1)^{mn}n e^{-\pi m n\tan\th}.
\eql{speffact}$$

The sum over $n$ produces
$$
W^{(3)}_{\rm sp}(\be')\equiv f_3'(0)=
\sum_{k=1}^\infty{1\over4k\sinh^2(k\be'/4)}-
\sum_{k=0}^\infty{1\over2(2k+1)\cosh^2\big((2k+1)\be'/8\big)},
\eql{speffact2}$$
where, for convenience, we have denoted $4\pi\tan\th$ by $\be'$. This
expression is useful numerically.

For small $l_3$, the exponent in (\peq{speffact}) can be written as 
$-m n \be/4$ with $\be$ {\it defined} by $\be=4\pi l_3$ and so,
dividing by $\be$ for normalisation purposes, we define the
corresponding spinor `free energy',
$$\Phi^{(3)}_{\rm sp}(\be)=-{1\over\be}W^{(3)}_{\rm sp}(\be')
\eql{therm2}$$

\begin{ignore}
For comparison, the fermion free energy on the product 
manifold $T\times\man$ is given by the standard formula
$$
F(\beta)=E+{1\over\beta}\sum_{m=1}^\infty {(-1)^m\over m}K^{1/2}(m\beta)
$$
where $E$ is the vacuum energy, and $K^{1/2}$ is the heat-kernel 
for the square root of the  Laplacian on $\man$. 
\end{ignore}

We will postpone discussion of formula (\peq{therm2}) until after 
the corresponding scalar case has been considered.

However before proceeding to this,
it is necessary to consider the other terms in (\peq{Planaq}) and 
(\peq{Planaqo}) which are
needed for the complete determination of the effective action as 
a function of the squashing.

In [\pref{dow11}], a further application of the Plana formula to the
$n$-sum in (\peq{Planaq}) revealed a series of  poles in $f(s)$ 
 at $s=3/2-m$, $m=1,2,\ldots$, with residues
$$
r_m=(-1)^{m+1}{2^{2m-2}\Ga(m-1/2)\over m!\Ga(1/2)}(l_3^2-1)^m 
l_3^{2m-2}\,B_{2m},
\eql{spres}$$
where $B_{2m}$ is a Bernoulli number (using the definition in Bateman
[\pref{Erdelyi}] \eg  In [\pref{dow11}] we used Lindel\"of's signs 
[\pref{Lindel}]).

Here we will employ the
Watson-Sommerfeld technique which is essentially equivalent to the
Plana one. It has been used by Shen and Sobczyk [\pref{SandS}] 
in the present context.

Completing the square, we have, 
$$
n^2+4\de^2(t^2-itn)=(n+iB)^2+A^2
\eql{compsq}$$ 
where
$$
A^2=4l_3^2(l_3^2-1)t^2 = \overline A^2t^2,\quad\quad
B=2(1-l_3^2)t=\overline B t\quad{\rm with}\,\, \overline B^2-\overline 
A^2=2\overline B.
$$
The signs
\etc are appropriate for the {\it prolate} case. For oblate $l_3<1$, 
we set $\overline A^2=-\overline C^2$.

We leave the $t$-integration until last and
just consider, for the oblate case first,
$$\eqalign{
\sum_{n=1}^\infty {n\over\big((n+iB)^2-C^2\big)^s}&={1\over 2i}
\int_L dz {z\over\big((z+iB)^2-C^2\big)^s}\cot\pi z\cr
&={1\over2i}\int_L d\ze {\ze-iB\over\big((\ze^2-C^2\big)^s}
\cot\big(\pi(\ze-iB)\big)\cr}
\eql{wats}$$
where $L$ is the anticlockwise contour surrounding the poles of the
cotangent at $\ze=n+iB$, $n=1,2,\ldots$. The $\ze$-plane has branch 
points at $-C$ and $+C$. For symmetry's sake the associated cuts are
arranged to run down the imaginary axis to (almost) the origin, and
thence along the real axis to $\pm C$. To make a choice, the $+C$-cut
comes in from $+i\infty$ and the $-C$ one from $-i\infty$. The 
contour $L$ is deformed to run down, just to the right of the imaginary
axis, then around the right-hand side of the $C$-cut and finally to 
skirt the imaginary axis down to $-i\infty$. We assume that $s$ is 
such as to ensure convergence. 

Taking the phases into account, the integral along the imaginary 
$\ze$-axis is
$$
-{1\over2}\int_0^\infty dy {y-B\over (y^2+C^2)^s} 
e^{-i\pi s}\coth\pi(y-B)-
{1\over2}\int_0^\infty dy {y+B\over (y^2+C^2)^s} 
e^{i\pi s}\coth\pi(y+B)
$$
and remembering that, according to (\peq{Planaqo}), we need 4 times the 
imaginary part of this expression we find,
$$
2\sin\pi s\bigg[
\int_0^\infty dy {y-B\over (y^2+C^2)^s}\coth\pi(y-B)-
\int_0^\infty dy {y+B\over (y^2+C^2)^s}\coth\pi(y+B)
\bigg]
$$

Now we make the split $\coth \pi x =1+2/(\exp 2\pi x-1)$. The `1'
part of this allows the integral to be done,
$$\eqalign{
2 \sin\pi s \bigg[-B C^{1-2s} {\sqrt\pi\Ga(s-1/2)\over\Ga(s)}-&
2\int_0^\infty dy {y+B\over (y^2+C^2)^s\big(\exp(2\pi(y+B))-1\big)}\cr
&+2\int_0^\infty dy {y-B\over (y^2+C^2)^s\big(\exp(2\pi(y-B))-1\big)}
\bigg],\cr}
\eql{wats2}$$ 
and it is straightforward to check that the first term in (\peq{wats2}) 
reproduces the poles in $f(s)$ at $s=3/2-m$, with residues (\peq{spres}),
after putting back the $t$-integration.

The integral over the right-hand part of the $C$-cut can likewise be 
reduced to the form, if it converges,
$$
4\sin\pi s\int_0^C dx {1\over (C^2-x^2)^s}{x\sinh(2\pi B)-
B\sin(2\pi x)\over\cosh(2\pi B)-\cos(2\pi x)}.
\eql{wats3}$$

From these expressions we can derive $f_2'(0)$,
$$
f_2'(0)=\!\int_0^\infty\!\!\! {4\pi t^2\,dt \over\exp(2\pi t)-1}
\bigg[\!\int_0^{\overline B}dy\,\! y\coth(\pi yt)\!
-\!\int_0^{\overline C}\!\!dx{x\sinh(2\pi\overline Bt)-\overline B
\sin(2\pi xt)\over\cosh(2\pi\overline Bt)-\cos(2\pi xt)}\bigg]
\eql{alto}$$
We note that, for $s$ a positive integer, the integral in (\peq{wats3}) 
does not converge so that this method is not convenient for these
values.

The prolate case can be treated in the same way and gives
$$
f_2'(0)={\overline A\,\overline B\over\pi^2}\ze(3)-4\pi\int_0^\infty
{t^2\,dt\over\exp(2\pi t)-1}\int_{\overline A+\overline B}^
{\overline A-\overline B}{y\,dy\over\exp (2\pi yt)-1}.  
\eql{altp}$$
In the near-round limit, $l_3\approx1$, $f_2'(0)$ goes like
$(1-l_3^2)/3$ in both cases. 

This approximation is most easily developed from the perturbation
expansion of the original sum in (\peq{Planaq}).
Since this provides a useful check, we give the first three terms 
obtained in this way.
Defining $l_3=\cosh\ga$ in the prolate and $l_3=\cos\phi$ in the oblate
case, we find
$$
f_2'(0)\approx-{1\over3}\ga^2-\bigg({1\over9}+{2\pi^2\over45}\bigg)\ga^4
-\bigg({2\over135}+{4\pi^2\over45}-{16\pi^4\over2835}\bigg)\ga^6
\eql{pertexp}$$
and the corresponding oblate form obtained by setting $\ga\to i\phi$.
Numerically, for $\ga=0.5$, (\peq{pertexp}) gives $-0.123$, in nice
agreement with numerical integration applied to (\peq{altp}).

To order $l_3^2$, the complete effective action from (\peq{detexpsn})
is
$$
W_{\rm sp}=-{1\over2\pi^2}\ze_R(3)+f_2'(0)+f_3'(0)+O(l_3^4)
\eql{sptot}$$
where $f_2'(0)$ is given by (\peq{altp}) in the prolate and by 
(\peq{alto}) 
in the oblate case. In the oblate case, $f_3'(0)$ is given
by (\peq{speffact}), otherwise it is zero. 

$W_{\rm sp}$, as given by (\peq{sptot}), is the quantity we 
concentrate on. While the expressions in this section are numerically
adequate, more suitable forms are developed in section 8.

For future reference, we note that when $\overline A$ and 
$\overline C$ are zero, but $\overline B\ne0$, both (\peq{alto}) and 
(\peq{altp}) give the same result, namely
just the first part of (\peq{alto}). One must, however, be cautious
with the $\overline C\to0$ limit in general, since the integrals in 
(\peq{wats2})
may not converge. If one did wish to pursue this further, then a 
different contour deformation would be called for. Furthermore, it
is possible to make a perturbation expansion in $C$, or in $A$, of the
original summation in (\peq{Planaq}) or (\peq{Planaqo}) the coefficients
of which depend on the (imaginary part of) the \zf\
$$
\sum_{n=0}^\infty{n\over(n+iB)^s}
$$
similar to the Hurwitz \zf\ with imaginary parameter. The general nature
of this expansion does not distinguish between prolate and oblate
deformations.
\section{4. \bf The scalar field}

We now turn to the massless scalar field. In the context of the
squashed three-sphere, this has earlier been considered by 
Critchley and Dowker [\pref{CrandD}], Shen and Sobczyk [\pref{SandS}], 
and more 
recently by Shtykov and Vassilevich [\pref{SandV}] who were concerned 
with the heat-kernel expansion for the Laplacian. We will use the 
scalar operator, $-\De+1/4$ for which the eigenvalues are 
$$
\la={1\over4l_3^2}
\bigg(n^2+4\de^2\big(q+{1\over2})(n-q-{1\over2}\big)\bigg)
\eql{seigenv}$$
with degeneracy, $n$. The $n$ label runs from $1$ to $\infty$, and
$q$ from $0$ to $n-1$.

We recall that the scalar curvature on $\widetilde S^3$ is
$$
R=2-{l_3^2\over2}
$$
which correctly gives $3/2$ when $l_3=1$ for the round three-sphere 
(of radius 2) and $2$ for the two-sphere at $l_3=0$. Our choice of
operator restricts to $-\De+R/6$ in the round case, and corresponds
to a conformally invariant wave equation in four dimensions. It also
reduces to $-\De+R/8$ on the unit two-sphere when $l_3=0$.

Other possible operators would be the minimal one, $-\De$,
and the one always conformally invariant in three dimensions $-\De+R/8$. 
It is
not difficult to accommodate these by an expansion as in the spinor case,
but since we are interested, for the moment, mostly in the leading 
high-temperature terms, the choice was dictated by the simplicity of the 
eigenvalues.

The \zf, $\sum \la^{-s}\equiv\ze_{sc}(s)$, is 
$$
\ze_{sc}(s)=(2l_3)^{2s}\sum_{n=1}^\infty\sum_{q=0}^{n-1}{n\over
\bigg(n^2+4\de^2\big(q+{1\over2})(n-q-{1\over2}\big)\bigg)^s}
\eql{effess}$$
and the effective action is $W_{\rm sc}=-\ze_{\rm sc}'(0)/2$.

The Plana summation formula is now applied, in the manner
of [\pref{dow11}] and discussed also in [\pref{SandS}]. We will consider
the oblate case. Again extra singularities appear
in the relevant band of the complex $q$-plane. Replacing $q$ by $z$ and
making the transformation of variable
$$
z={n-1\over2}+i{n\over2}\ze\tan\th
$$
the denominator becomes
$$
\big(l_3^2n^2(1-\ze^2)\big)^s
$$
showing the branch points at $\ze=\pm1$, as before. 

In this scalar case we use the alternative choice of summation 
contour, the band in the $z$ plane being defined by $-1/2\le \Real z\le
(n-1/2)$ (rather than $0$ to $n-1$). Then the full Plana formula,
applied just to the $q$ summation, is
$$\eqalign{
\ze_{sc}(s)=&2^{2s}\tan\th\,\ze_R(2s-2)\int_0^{\cot\th}{dy\over
(1+y^2)^s}\,+\cr
&2i(2l_3)^{2s}\int_0^\infty {dt\over \exp(2\pi t)+1}\bigg\{\sum_{n=1}
^\infty{n\over\big(n^2+4\de^2(t^2-itn)\big)^s}-(t\to -t)\bigg\}\,-\cr
&i2^{2s}\tan\th\sum_{n=1}^\infty {1\over n^{2s-2}}
\int_{C} {d\ze\over \big((-1)^n
\exp(n\pi\ze\tan\th)+1\big)}{1\over\big(1-\ze^2\big)^s},\cr}
\eql{Planaqs}$$

Look at the last term first and again combine the upper and lower
parts of the clockwise loop $C$ to give 
$$
-2^{2s+1}\sin(\pi s)\tan\th\sum_{n=1}^\infty{1\over n^{2s-2}}
\int_1^\infty {d\ze\over \big((-1)^n
\exp(n\pi\ze\tan\th)+1\big)}{1\over\big(\ze^2-1\big)^s},
\eql{xtraint3}$$
when the integration converges at $\ze=1$. 

The contribution to $-\ze_{sc}'(0)/2$ from this term is then
$$
W_{\rm sc}^{(3)}=\pi\tan\th\sum_{n=1}^\infty n^2\int_1^\infty 
{d\ze\over (-1)^n\exp(n\pi\ze\tan\th)+1}
\eql{diff0s}$$
which can be written as
$$
W_{\rm sc}^{(3)}(\be')=\sum_{m=1}^\infty {1\over m} \sum_{n=1}^
\infty (-1)^{m(n+1)}\,n\,e^{-nm\be'/4},
\eql{therms1}$$
and sums to
$$
W^{(3)}_{\rm sc}(\be')=
\sum_{k=1}^\infty{1\over8k\sinh^2(k\be'/4)}+
\sum_{k=0}^\infty{1\over4(2k+1)\cosh^2\big((2k+1)\be'/8\big)}.
\eql{speffact3}$$

The corresponding contribution to 
the free energy $\Phi^{(3)}_{\rm sc}(\be)={1\over\be}\ze_{sc}'(0)$ is
$$
\Phi^{(3)}_{\rm sc}(\be)=-{1\over\be}W^{(3)}_{\rm sc}(\be')
\eql{therms2}$$

When $l_3$ is zero, the metric (\peq{3metric}) reduces to that of the
unit two-sphere.
The scalar eigenvalues of $-\De+1/4$ on the unit two-sphere are $n^2/4$
with odd $n$. The degeneracy is $n$ and so the standard finite 
temperature correction to the bosonic free energy reads
$$
F'_{\rm sc}(\beta)=-{1\over\beta}\sum_{m=1}^\infty {1\over m}
\sum_{n=1,3,\ldots}^\infty n\, e^{-m n \be/2}.
\eql{freeens}$$

The eigenvalues of the massless 
{\it Dirac} operator the unit $S^2$  are $\pm n$, $n=1,\ldots,
\infty$ with degeneracy $2n$. So, (including positive and negative modes)
$$
F'_{\rm sp}(\beta)={2\over\beta}\sum_{m=1}^\infty {(-1)^m\over m}
\sum_{n=2,4,\ldots}^\infty n\, e^{-m n \be/2}.
\eql{freeen4}$$

The effective
action densities on the squashed three-sphere, in the highly oblate
direction seem to possess both boson and fermion characteristics. 

Split the sum in (\peq{therms1}) into even and odd $n$,
$$
\sum_{m=1}^\infty {1\over m} \sum_{n=1,3,\ldots}^\infty  
n\, e^{-nm\be'/4}+
\sum_{m=1}^\infty {(-1)^m\over m} \sum_{n=2,4,\ldots}^\infty  
n\, e^{-nm\be'/4}
\eql{therms3}$$ 
so we can write, at least formally,
$$
\Phi^{(3)}_{\rm sc}(\be)={\be'\over\be}\bigg({1\over2}
F_{\rm sc}'(\be'/2)-{1\over4}F_{\rm sp}'(\be'/2)\bigg).
\eql{comb1}$$
Hence the small $\be$ behaviour can be determined from known
results on the two-sphere.

The standard expressions for the Weyl terms, for thermal behaviour on
a two-manifold, here the two-sphere, are
$$
F'_{\rm sc}(\be)\sim -{\ze_R(3)\over2\pi\be^3}C_0^{\rm sc}=-{2\ze_R(3)
\over\be^3}\eql{leadsc}$$
and
$$
F'_{\rm sp}(\be)\sim -{3\ze_R(3)\over8\pi\be^3}C_0^{\rm sp} 
=-{3\ze_R(3)\over\be^3}
\eql{leadsp}$$
where $C_0^{\rm sc}=|\man|$ and $C_0^{\rm sp}=|\man|\,\Tr{\bf 1}$. The 
coefficients of the $C_0$'s and indeed all {\it formal} expressions, 
satisfy the functional relation
$$
Q'_{\rm sp}(\be)=Q'_{\rm sc}(\be)-2Q'_{\rm sc}(2\be).
\eql{reln}$$

From (\peq{comb1}) the leading terms are
$$
\Phi_{\rm sc}^{(3)}(\be)\sim-{2\ze_R(3)\over\be^3}
+{1\over16\pi^2}{\ze_R(3)\over\be}
+{1\over12\be}\ln\be+{1\over\be}\ze_R'(-1).
\eql{3asymps}$$

This result can be obtained more directly from the summed form 
(\peq{speffact3}), which differs from the spinor form, 
(\peq{speffact2}), just in the sign of the second term, and a factor of
two. Moreover, this 
term has a logarithmic asymptotic dependence on $\be$ as $\be\to0$,
which is expected since the $C_1$ heat-kernel coefficient on the
two-sphere is non-zero. The corresponding sub-leading terms can be seen
in the last two terms in (\peq{3asymps}). The scaling of the $\ln\be$ 
is provided by the size of the two-sphere (for which there {\it is} a 
conformal anomaly). The second term on the right-hand side of 
(\peq{3asymps})
is a simple consequence of the difference between `effective' inverse 
temperature, $\be'$, and the `true' one, $\be$.

For the record, we give the summed forms of the scalar and spinor
free energies on the two-sphere,
$$
F'_{\rm sc}(\be)=-{1\over2\be}\sum_{m=1}^\infty 
{\cosh(m\be/2)\over m\sinh^2(m\be/2)}
\eql{s2scfe}$$
$$
F'_{\rm sp}(\be)={1\over\be}\sum_{m=1}^\infty 
{(-1)^m\over m\sinh^2(m\be/2)}.
\eql{s2spfe}$$

We must now turn to the $n$-summation in (\peq{Planaqs}). The
calculation is much the same as for the spinor field.

The prolate form obtained by the Watson-Sommerfeld method is,
$$
W_{\rm sc}^{(2)}=
{3\overline A\,\overline B\over8\pi^2}\ze_R(3)-2\pi\int_0^\infty
{t^2\,dt\over\exp(2\pi t)+1}\int_{\overline A+\overline B}^{\overline 
A-\overline B}{y\,dy\over\exp (2\pi yt)-1}.
\eql{saltp}$$
This is handy if we want the extreme prolate (large $l_3$) limit.
From the first term, the leading behaviour of the entire
scalar effective action is 
$$
W_{\rm sc}\approx W_{\rm sc}^{(2)}\sim -{3l_3^4\over2\pi^2}\ze_R(3),
\quad l_3\to\infty
\eql{hi}$$
showing a sort of duality.

The oblate expression is found to be,
$$
W_{\rm sc}^{(2)}=\!\int_0^\infty\!\!\! {2\pi t^2\,dt \over\exp(2\pi t)+1}
\bigg[\!\int_0^{\overline B}\!\!\!dy\, y\coth(\pi yt)\!
-\!\int_0^{\overline C}\!\!\!dx{x\sinh(2\pi\overline Bt)-\overline B
\sin(2\pi xt)\over\cosh(2\pi\overline Bt)-\cos(2\pi xt)}\bigg],
\eql{salto}$$
and the total scalar free energy, from (\peq{Planaqs}), is
$$
W_{\rm sc}={\ze_R(3)\over4\pi^2}+W_{\rm sc}^{(2)}+W_{\rm sc}^{(3)}.
\eql{sctots}$$

For prolate deformations, $W_{\rm sc}^{(2)}$ is given by (\peq{saltp})
and for oblate by (\peq{salto}). There is no $W_{\rm sc}^{(3)}$ for the
prolate case and it equals (\peq{speffact3}) in the oblate.

The corresponding `free energy' in both cases is defined to be 
$\Phi_{\rm sc}(\be)=-W_{\rm sc}/\be$ and from (\peq{sctots}) 
we see that $\Phi_{\rm sc}(\be)\sim 
\Phi^{(3)}_{\rm sc}(\be)$ so that from (\peq{3asymps}) this quantity 
has the same high temperature behaviour as on $S^2$. This statement is 
expanded in section 9.

Again we note that, in the $\overline A=0$ and $\overline C=0$ limits,
both (\peq{saltp}) and (\peq{salto}) reduce to just the first
term of (\peq{salto}).

\section {\bf 5. Spinors revisited.}

We now return to (\peq{therm2}) for the spin-half (Pauli) free 
energy on the squashed three-sphere and again decompose the sum
into even and odd pieces,
$$
\Phi^{(3)}_{\rm sp}(\be)=
-{2\over\be}\sum_{m=1}^\infty {1\over m} 
\sum_{n=2,4,\ldots}^\infty n\,e^{-m n\be'/4}-
{2\over\be}\sum_{m=1}^\infty {(-1)^m\over m} 
\sum_{n=1,3,\ldots}^\infty n\, e^{-m n\be'/4}
\eql{therm5}$$
which presents a sort of twisted situation with thermal fermions
being associated with scalar modes on $S^2$ and {\it vice versa}.

Let us rewrite this expression as
$$
\Phi^{(3)}_{\rm sp}(\be)={\be'\over\be}\bigg({1\over2}\widetilde 
F_{\rm sp}(\be'/2)-\widetilde F_{\rm sc}(\be'/2)\bigg)
\eql{comb2}$$
to be compared with (\peq{comb1}).

To obtain the leading behaviour we can make use of the fact that the
functional relation (\peq{reln}) takes the twisted form,
$$
\widetilde Q'_{\rm sc}(\be)=\widetilde Q'_{\rm sp}(\be)-
2\widetilde Q'_{\rm sp}(2\be).
\eql{reln2}$$

The coefficients of the $C_0$'s in (\peq{leadsc}) and 
(\peq{leadsp}) will then be switched \ie
$$
\widetilde F'_{\rm sp}(\be)\sim -{\ze_R(3)\over2\pi\be^3}C_0^{\rm sp}
=-{4\ze_R(3)\over\be^3}
\eql{leadsct}$$
and
$$
\widetilde F'_{\rm sc}(\be)\sim -{3\ze_R(3)\over8\pi\be^3}C_0^{\rm sc} 
=-{3\ze_R(3)\over2\be^3}
\eql{leadspt}$$
then  from (\peq{comb2})
$$
\Phi^{(3)}_{\rm sp}(\be)\sim-{4\ze_R(3)\over\be^3}
+{1\over8\pi^2}{\ze_R(3)\over\be}-
{1\over6\be}\ln\be-{2\over\be}\ze_R'(-1).
\eql{spasymp}$$
Again, this follows also from the summed form (\peq{speffact2}).

Once more we see that the leading behaviour on the extreme oblate
three-sphere corresponds exactly to the high temperature form
on the two-sphere, but not so the sub-leading terms.

\section{\bf 6.  Poles, residues and coefficients.} 

In [\pref{dow11}] we determined the analytic structure of the spinor
zeta function, and thence the coefficients in the 
heat-kernel
expansion. The same procedure can be pursued of course in the scalar 
case and just needs the complete pole structure of $\ze_{\rm sc}(s)$. 
The intermediate Plana form
(\peq{Planaqs}) shows a pole at $s=3/2$ coming from the Riemann
\zf. This is the `volume' pole. It has a residue $4l_3$ which
agrees with the general value, 
$$
{C_m\over(4\pi)^{d/2}\Ga\big(d/2-m\big)}
$$
for the residue at $s=d/2-m$ ($m=0,1,\ldots,\,\ne d/2)$ in a 
$d$-dimensional manifold, $\man$, if we recall the Weyl result, 
$C_0=|\man|$. For our squashed three-sphere, 
$|\widetilde S^3|=16\pi^2 l_3$.

To obtain all the poles, we can proceed 
as described in section 2 the only differences being the change in sign
and the bosonic factor. In the scalar case
$\ze_{\rm sc}(s)$ has poles at $s=3/2-m$, $m=1,2,\ldots$ with residues
$$
R_m=(2l_3)^{3-2m}(2^{1-2m}-1)r_m
$$
in terms of the spinor residues (\peq{spres}). Hence the heat-kernel
coefficients are 
$$
C_m= |\widetilde S^3|(2^{1-2m}-1)(l_3^2-1)^m\,{B_{2m}\over m!}.
\eql{hkcoeffs1}$$

Removing the volume factor, and taking the $l_3\to0$ limit yields
Mullholland's results for the unit two-sphere scalar coefficients,
$$
\lim_{l_3\to0}{C_m\over|\widetilde S^3|}=(-1)^m(2^{1-2m}-1)\,{B_{2m}
\over m!}.
$$
The spin-half case is similarly treated. 

Shtykov and Vassilevich [\pref{SandV}] have computed the coefficients 
on the deformed
unit three-sphere for the minimal operator, $-\De$, To get these 
coefficients one can simply multiply the above expansion by the
expansion of $\exp(t/4)$ in the usual way. For example 
$$
C_1^{\rm minimal}=|\widetilde S^3|{4-l_3^2\over 12}=|\widetilde S^3| 
{R\over6}
$$
as required.
\section{\bf 8. An alternative summation and a puzzling coincidence.}

It is always a good idea to pursue different paths to any required
quantity. Apart from inspiring confidence, or otherwise, in the final 
answer, it often reveals unexpected subleties.

Instead of applying the Watson-Sommerfeld method to the $n$-summations
in (\peq{Planaq}), (\peq{Planaqo}) and (\peq{Planaqs}), we can proceed
as Okada does, 
[\pref{Okada}], and employ the method first used by Minakshisundaram
when discussing the \zfs\ on spheres and later, more
extensively, by Candelas and Weinberg [\pref{CandW}] for the same
purpose. The idea is to rewrite the summand using the Laplace 
transform
$$
{1\over\big((n+iB)^2+A^2\big)^s}={\sqrt\pi(2A)^{1/2-s}
\over\Ga(s)}\int_0^\infty dz\,e^{-(n+iB)z}
z^{s-1}\,J_{s-1/2}(Az)
\eql{ltrans}$$
in order to effect the $n$-summation.

For oblate $l_3<1$, 
$A$ becomes imaginary ($\overline A^2=-\overline C^2$) and the Bessel 
function becomes a modified one, $I_{s-1/2}$. However there is an 
obstruction to the immediate application of a formula such as 
(\peq{ltrans}) in that, for $t>1/\overline C$, $C$ will be larger than
at least one $n$, violating the conditions of the identity. Our 
attitude {\it in this section} is to take the prolate expression as far 
as possible, and 
then to continue in $l_3^2$. The form of the eigenvalues shows that
the singularities of the \zf, as a function of $l_3^2$, all lie
on the negative real axis and therefore should not prevent the
continuation through $l_3^2=1$.

Inserting (\peq{ltrans}) into the middle term of (\peq{Planaq})
allows the $n$-sum to be done and we find
$$
-{\sqrt\pi\,2^{1/2-s}\over\Ga(s)}\int_0^\infty dz {z\sin Bz\over
\sinh^2(z/2)}\, z^{2s-2} (Az)^{1/2-s}\,J_{s-1/2}(Az)
\eql{nsum1}$$
omitting the overall $t$-integral for the time being. At this point,
following Candelas and Weinberg, the $z$ integral is replaced by
a contour one in the cut, complex $z$-plane 
(this could have been done earlier) to get for (\peq{nsum1})
$$
-{\sqrt\pi\,2^{1/2-s}\over(1+\exp\big(2i\pi s)\big)\Ga(s)}\int_\cac dz 
{z\sin Bz\over\sinh^2(z/2)} z^{2s-2} (Az)^{1/2-s}\,J_{s-1/2}(Az)
\eql{nsum2}$$
where $\cac$ can be taken as, say, $z=x+i Y$ ($-\infty\le x \le \infty$). 
The
choice of the constant $Y$ depends on any singularities possessed by the
integrand. In this case, there are poles at $z=2pi\pi$ ($p=\pm1,\pm2,
\ldots$) and so $Y$ should be less than $2\pi$.

At this point we can again check the form of the residues (\peq{spres}) 
at the
poles of $f(s)$, which, in the present representation, occur at the 
zeros of $1+\exp(2i\pi s)$, \ie at half odd integer $s$. For $s>1/2$, 
the original form, (\peq{nsum1}), is convergent so the only possible
poles are at $s=3/2-m$, $m=1,2,\ldots$. For these values of $s$
there is no cut in the $z$-plane and, also, the integrand is an odd 
function of $z$. Hence, we can add the expression for the reversed
contour, $-\cac$ ($z=x-iY$, $x$ running from $+\infty$ to $-\infty$)
and divide by two. These two contours combine to give a
clockwise contour around the pole at the origin and the
integral is simply evaluated by residues. 

For (\peq{nsum2}) we obtain the value, putting back the $t$-integration,
$$
-{2(-1)^{m+1}\sqrt\pi\over\Ga(3/2-m)(m-1)!} \overline A^{2m-2}
\overline B\int_0^\infty dt{t^{2m-1}\over \exp(2\pi t)-1}
$$
and we 
regain (\peq{spres}) which expression can be continued without ambiguity
into the oblate region. Of course, being geometrical, the final result
must be valid in both the prolate and oblate cases.

Our main concern is with the derivative at zero, $f'(0)$.
In (\peq{nsum2}) one must differentiate the $1/\Ga(s)$ factor and set
$s=0$ in the rest to obtain a non-zero answer. We find for the 
corresponding contribution to $f'(0)$,
$$
f_2'(0)= -\int_0^\infty {dt\over\exp(2\pi t)-1}
\int_\cac {dz\over z\sinh^2(z/2)}\sin \overline Btz\, 
\cos \overline Atz 
\eql{inter1}$$ 
and now the $t$-integral can be done (interchanging the limiting
processes) using the standard formula
$$
\int_0^\infty\,dt\,{\sin at\over\exp(2\pi t)-1)}={1\over4}\coth(a/2)-
{1\over2a},\quad |\Imag a| < 2\pi,
$$
to produce
$$\eqalign{
f_2'(0)=&-{1\over4}\int_\cac{dz\over z\sinh^2(z/2)}\bigg[
\sumdasht{p=-\infty}{\infty}
\bigg({1\over\big((\overline B+\overline A)z-2\pi ip\big)}+
{1\over\big((\overline B-\overline A)z-2\pi ip\big)}\bigg)\bigg]\cr
=& {1\over4}\int_\cac{dz\over z^2\sinh^2(z/2)}\bigg(1-{z\sinh \overline Bz
\over \cosh \overline Bz-\cosh \overline Az}\bigg)\cr
=&{1\over2\pi^2}\ze_R(3)-
{1\over4}\int_\cac{dz\over z\sinh^2(z/2)}\bigg({ \sinh \overline Bz
\over \cosh \overline Bz-\cosh \overline Az}\bigg)\cr}
\eql{fdash}$$
The integrand is even in $z$ ensuring that the integral is real.
Convergence at large $|z|$ is also assured. We also note that the 
$\ze_R(3)$ term cancels against the contribution to $f'(0)$ from the 
first term in (\peq{sptot}).

The price to be paid for performing the $t$-integration is a further
constriction of the contour $\cac$, for which, in (\peq{nsum2}), 
$Y$ lies between $0$ and $2\pi$. The condition on the validity of the 
$t$-integration means that now $0<Y<2\pi/|B\pm A|$, or, if we set 
$l_3=\cosh\ga$, $0<Y<\pi(\coth\ga\pm1)$.

The term in brackets in (\peq{fdash}) presents poles at
$$
z_p^{(\pm)}={2i\pi p\over \overline B\pm \overline A}
=i p \pi\big(1\pm {l_3\over\sqrt{l_3^2-1}}\big)\equiv 
ip\pi(1\pm\coth\ga),\quad p=\pm1,\pm2,\ldots.
\eql{npolesp}$$

Starting from the prolate side, all the plus sign poles, $z_p^{(+)}$,
 $p>0$, in 
(\peq{npolesp}) lie above the lowest existing pole at $2\pi i$ and 
cause no
problems. However the minus sign, $p=-1$ pole, $z_{-1}^{(-)}$, falls 
below this pole when $l_3>3/(2\sqrt2)\approx1.06066$ and approaches the
origin as $l_3$ increases, as do all the other minus sign poles, 
symmetrically with the sign of $p$. 
The contour $\cac$ has to be adjusted to lie below this pole. This is
what the above condition means.

Just as a check, we have numerically evaluated (\peq{inter1}) and
(\peq{fdash}). Typically, for $\ga=0.5$, ($l_3=1.1276$) we find
that (\peq{inter1}) gives $f_2'(0)\approx -0.123$, for 
values, $Y=3$ and $Y=4$, which straddle the pole
at $z=i\pi(\coth\ga-1)\approx i3.6565$, the order of integration
making no difference. In contrast, (\peq{fdash}) gives $-0.123$
if $Y$ is below this pole, but $0.143$ if above. The
difference is accounted for by the pole residue, as has been numerically
checked. 

As the oblate case is approached, $l_3\downarrow1$, the plus sign poles,
$z_p^{(+)}$, with positive $p$ run away up the imaginary axis to 
$i\infty$, as do
the minus sign poles, $z_p^{(-)}$, with {\it negative} $p$. Conversely, 
the 
plus sign poles with negative $p$, together with the minus sign poles with 
{\it positive} $p$, run off to $-i\infty$. 

As $l_3$ passes into the oblate r\'egime, the poles reappear at the 
complex positions
$$
z_p^{(\pm)}=p\pi(i\pm \tan\th)
\eql{obpoles}$$ 
We should follow them round carefully. It is apparent that some
prolate poles with positive (negative) $p$, disappear with negative 
(positive) imaginary  parts, only to reappear as oblate poles 
{\it with imaginary parts of opposite sign}. They have thus crossed the
contour $\cac$ (at complex infinity) and, if we want to leave $\cac$ 
unchanged, it is necessary to include the corresponding residue 
contributions. These yield, choosing $\sqrt{l_3^2-1}=i\sqrt{1-l_3^2}$,
$$
{1\over2}\sum_{p=1}^\infty{1\over p\sinh^2 z_p^{(-)}}.
$$
Symmetry under $A\to-A$ (or $C\to-C$) is maintained automatically
because $\sinh^2 z_p^{(-)}=\sinh^2 z_p^{(+)}$, in tune with the fact
that the branch point in (\peq{npolesp}) at $l_3^2=1$ is purely
artificial. 

The complete oblate contribution $f_2'(0)$ is then
$$\eqalign{
f_2'&(0)= {1\over2\pi^2}\ze_R(3)-{1\over4}\int_\cac
{dz\over z\sinh^2(z/2)}\bigg({\sinh \overline Bz
\over \cosh \overline Bz-\cos \overline Cz}\bigg)+\cr
&{1\over2}\bigg(\sum_{m=1}^\infty{1\over2m\sinh^2(m\pi\tan\th)}-
\sum_{m=0}^\infty{1\over(2m+1)\cosh^2\big((2m+1)\pi\tan(\th)/2\big)}
\bigg).\cr}
\eql{fdasho}$$

Other values of interest are $f(n)$, with $n$ a positive integer. These
are needed in the expansion (\peq{detexpsn}). Looking at (\peq{Planaqo})
the first two terms are easily evaluated and the last one vanishes,
as shown earlier. The remaining one, $f_2(s)$, is under present
investigation and we find from (\peq{nsum2}),
$$
f_2(n+1)= -{4\sqrt\pi\over n!}\!\int_0^\infty\!\!\!{dt\over\exp(2\pi t)
-1}\int_0^\infty\!\!\! dz {\sin\overline Btz\over
\sinh^2(z/2)} z^{2n+1} (\overline Atz)^{-n-1/2}\,J_{n+1/2}(\overline 
Atz)
\eql{fint}$$

This time the $z$ integration can be done. As an example
consider $f_2(1)$
$$
f_2(1)=-4\sqrt2\int_0^\infty{t^{-1}dt\over\exp(2\pi t)-1}\int_0^\infty 
dz{\sin(\overline Btz)\sin(\overline Atz)\over \overline A\sinh^2z/2}
$$
and use the standard integral
$$
\int_0^\infty dx\,{\sin ax\sin bx\over\sinh^2x/2}=2\pi
{b\sinh(2a\pi)-a\sinh(2b\pi)\over\cosh(2a\pi)-\cosh(2b\pi)}
\eql{stint2}$$
to leave a single numerical quadrature. In the oblate case 
($\overline A\to i\overline C$), the integral in
(\peq{stint2}) always converges. 

The other values can be reduced similarly to involve integrals 
obtained by repeated differentiation of (\peq{stint2}).

For these values of $s>3/2$, one can also calculate 
$f(s)$ by direct summation which, purely numerically, is probably
more convenient.

For scalar fields, the evaluation of the $n$-summation term in 
(\peq{Planaqs}) is precisely the same as decribed above for the spin-half
case. The only changes are the overall sign of the term and the 
`bosonic' factor $1/(\exp(2\pi t)+1)$, which results in some algebraic
differences. In place of (\peq{inter1}) we have (remember $W=-\ze'(0)/2$
here)
$$
W_{\rm sc}^{(2)}=- {1\over2}\int_0^\infty {dt\over\exp(2\pi t)+1}
\int_\cac {dz\over z\sinh^2(z/2)}\sin \overline Btz\, 
\cos \overline Atz 
\eql{inter2}$$
and this time we use the integral
$$
\int_0^\infty\,dt\,{\sin at\over\exp(2\pi t)+1)}=
{1\over2a}-{1\over4}{\rm cosech}\,(a/2),\quad |\Imag a|<2\pi,
$$
to give in the prolate case,
$$\eqalign{
W_{\rm sc}^{(2)}=&{1\over8}\int_\cac{dz\over z\sinh^2(z/2)}
\sumdasht{p=-\infty}{\infty}
\bigg({(-1)^p\over\big((\overline B+\overline A)z-2\pi ip\big)}-
{(-1)^p\over\big((\overline B-\overline A)z-2\pi ip\big)}\bigg)\cr
=&-{1\over4\pi^2}\ze_R(3) +{1\over4}\int_\cac{dz\over z\sinh^2(z/2)}
{\sinh (\overline Bz/2)\cosh(\overline Az/2)
\over \cosh \overline Bz-\cosh \overline Az}\cr}
\eql{fdashs}$$
with the same restriction on the contour, $\cac$.

The oblate expression is found to be
$$\eqalign{
W_{\rm sc}^{(2)}&=-{1\over4\pi^2}\ze_R(3)+ {1\over4}\int_\cac{dz\over 
z\sinh^2(z/2)}{\sinh (\overline Bz/2)\cos(\overline Cz/2)
\over \cosh\overline Bz-\cos\overline Cz}-\cr
&{1\over4}\bigg(\sum_{m=1}^\infty{1\over2m\sinh^2(m\pi\tan\th)}+
\sum_{m=0}^\infty{1\over(2m+1)\cosh^2\big((2m+1)\pi\tan(\th)/2\big)}
\bigg)\cr}
\eql{fdashso}$$
When this is substituted into the total expression (\peq{sctots}) 
we again note the cancellation of the $\ze_R(3)$ terms.

There appears to be no ambiguity in 
the prolate calculation, and we have precise numerical agreement
between (\peq{fdash}) and (\peq{altp}) for spinors, and between
(\peq{fdashs}) and (\peq{saltp}) for scalars.

However, extra pole terms seem to arise when the Bessel technique is 
`continued' from the prolate to the oblate r\'egime, as
exhibited in (\peq{fdasho}) and (\peq{fdashso}). These terms do not
affect the perturbation expansion and indeed the first lines of 
(\peq{fdashso}) and (\peq{fdasho}) equal the whole of (\peq{salto}) 
and (\peq{alto}) respectively. Since they involve just a single
quadrature, they at least provide better numerical alternatives. 

The puzzle is, partly, that these extra terms precisely  cancel the 
oblate contribution, $W^{(3)}_{\rm sc}$ (\peq{speffact3}), in the 
scalar case, and double this up in the spinor case. This is a curious 
coincidence that we cannot explain. 

If one did believe the forms derived in this section, then there is no
divergence in the {\it total} effective action as $l_3\to0$, for 
scalar fields. This would seem to run counter to
a general expectation, derived from experience
with the Selberg-Chowla formula, and also shows an unbelievable
difference between scalars and spinors. 

It is clear, however, that the results on the $\overline C=0$ case
based on a perturbation expansion of the original sum form about
$\overline C=0$ (or about $\overline A=0$), discussed in the earlier 
sections, show that there is no
divergence in $W^{(2)}$ as $l_3\to0$ and hence that the results
presented in the previous sections of this paper are the proper ones.
We therefore discard the second lines of (\peq{fdashso}) and 
(\peq{fdasho}). Doing this allows one to find the $l_3\to0$ limits of 
$W^{(2)}$ in closed form,
$$
W_{\rm sc}^{(3)}\bigg|_{l_3=0}+{1\over4\pi^2}\ze_R(3)=
{1\over16}\int_\cac {dz\over z\sinh^3z\cosh z}={1\over4}\ln2-{1\over8
\pi^2}\ze_R(3) 
\eql{sc2}$$
and
$$
W_{\rm sp}^{(3)}\bigg|_{l_3=0}-{1\over2\pi^2}\ze_R(3)=
-{1\over8}\int_\cac {\cosh 2z\, dz\over z\sinh^3z\,\cosh z}
={1\over2}\ln2+{1\over4
\pi^2}\ze_R(3)
\eql{sp2}$$
where the contour has been pushed upwards through the poles of the
integrand and the residues summed.

\section{\bf 9. Preliminary comparison with the bulk results.}

We are now in a position to gather our results together and give
the total high temperature forms of the free energy, defined by
$\Phi=-W/\be$.

For the scalar field, from (\peq{sctots}), (\peq{sc2}) and 
(\peq{3asymps}),
$$
\Phi_{\rm sc}(\be)\sim -{2\ze_R(3)\over\be^3}+{1\over12\be}\ln\be
+{1\over\be}\bigg({3\over16\pi^2}\ze_R(3)-{1\over4}\ln2+\ze'(-1)\bigg),
\eql{totalsc}$$
while, for the spinor field from (\peq{sptot}), (\peq{sp2}) and 
(\peq{spasymp}),
$$
\Phi_{\rm sp}(\be)\sim -{4\ze_R(3)\over\be^3}-{1\over6\be}\ln\be
-{1\over\be}\bigg({1\over8\pi^2}\ze_R(3)+{1\over2}\ln2+2\ze'(-1)\bigg).
\eql{totalsp}$$

According to [\pref{HHP}] and [\pref{CEJM}], the Taub-Bolt-AdS
bulk, four-dimensional entropy goes like $1/\be^2$ at high
temperatures. This behaviour agrees, of course, with our
expressions for the free energy, or action, with which we prefer to 
work.

Extracting the leading terms from [\pref{HHP}]
$$
\be\Phi\sim -\al\bigg({1\over\be^2}
-{9\over8\pi^2}+{27(k+2)^2\over1024\pi^4}\be^2\bigg)  
\eql{ADS}$$
where $\al$ is a constant which does not concern us here. The 
identification parameter $k$ has been included and should be set equal 
to unity to conform to the assumptions of the present paper. In this 
case, (\peq{ADS}) agrees with the calculations in [\pref{CEJM}].

We notice immediately the absence of any
logarithmic or transcendental terms. These can be eliminated from our
expressions by choosing a combination of two scalars plus one spinor, 
yielding the behaviour
$$
\be\Phi\bigg|_{\rm two\,\, scalars\,\,+\,\,one\,\, spinor}\sim
-{8\ze_R(3)}\bigg({1\over\be^2}-{1\over108\pi^2}\bigg)
\eql{2s1s}$$
which is the best that can be done with the fields available.
\section{\bf 10. Conclusion.}

We have presented formulae for the determinants of spinor and
scalar fields on the squashed three-sphere and have determined
explicitly their leading $1/\be^2$ behaviours which exactly correspond
to those on $S^2\times S^1$. The non-leading behaviours have
also been found where the twisted nature of $\widetilde S^3$ shows up.
The scalar expression is a mixture of thermal scalar and spinor on 
$S^2$ while the spinor form combines twisted scalar and spinor,
\ie fields with the `wrong' thermal periodicity.

The next step in the calculation, which will be exposed at a later 
date, is to
place $k$ identifications on the $\psi$-circle. Since the resulting
manifold is more or less locally unchanged, one would expect the 
highest terms to remain the same, depending, as they do, on the 
$S^2$ geometry. This is born out in (\peq{ADS}) and (\peq{2s1s}), 
and is a comment made also in [\pref{CEJM}].

The extreme {\it prolate} limit presents certain technical difficulties 
which have still to be addressed. The action for the branch of the bulk 
action that leads to (\peq{ADS}) tends to a constant as $l_3\to\infty$
while the other branch diverges like $l_3^4$, in agreement with (\peq{hi})
for the scalar field. For the spinor field, one needs to go beyond
the expansion in $l_3$ and to allow for sign changes in the spectrum. 
For the scalar field we should also compute for the minimal Laplacian. 

It should be noted that our $W$ corresponds to $-I$ of [\pref{HHP,CEJM}].
$I$ is the difference between the actions of Taub-Bolt-AdS and 
Taub-Nut-AdS and is only defined for certain regions of $l_3$. There 
appears to be no corresponding limitation in the present calculation since
the round case, $l_3=1$, is always accessible, whatever the operator. 

Finally we remark that it is necessary to consider vector fields, for 
which the eigenvalues are also available, [\pref{dow11,Gibb}]. 

\section{\bf Acknowledgments}
I would like to thank Stephen Hawking for suggesting that this
calculation was worth doing and therefore for reawakening my interest
in the squashed three-sphere. Don Page also made useful comments.
Thanks are also due to the Theory Group at Imperial College, where this
work was performed and also to the EPSRC for providing support under
Grant No GR/M08714.

\section{\bf References}
\vskip 5truept
\begin{putreferences}
\ref{APS}{Atiyah,M.F., V.K.Patodi and I.M.Singer: Spectral asymmetry and 
Riemannian geometry \mpcps{77}{75}{43}.}
\ref{AandT}{Awada,M.A. and D.J.Toms: Induced gravitational and gauge-field 
actions from quantised matter fields in non-abelian Kaluza-Klein thory 
\np{245}{84}{161}.}
\ref{BandI}{Baacke,J. and Y.Igarishi: Casimir energy of confined massive 
quarks \prD{27}{83}{460}.}
\ref{Barnesa}{Barnes,E.W.: On the Theory of the multiple Gamma function 
{\it Trans. Camb. Phil. Soc.} {\bf 19} (1903) 374.}
\ref{Barnesb}{Barnes,E.W.: On the asymptotic expansion of integral 
functions of multiple linear sequence, {\it Trans. Camb. Phil. 
Soc.} {\bf 19} (1903) 426.}
\ref{Barv}{Barvinsky,A.O. Yu.A.Kamenshchik and I.P.Karmazin: One-loop 
quantum cosmology \aop {219}{92}{201}.}
\ref{BandM}{Beers,B.L. and Millman, R.S. :The spectra of the 
Laplace-Beltrami
operator on compact, semisimple Lie groups. \ajm{99}{1975}{801-807}.}
\ref{BandH}{Bender,C.M. and P.Hays: Zero point energy of fields in a 
confined volume \prD{14}{76}{2622}.}
\ref{BBG}{Bla\v zi\' c,N., Bokan,N. and Gilkey,P.B.: Spectral geometry 
of the 
form valued Laplacian for manifolds with boundary \ijpam{23}{92}{103-120}}
\ref{BEK}{Bordag,M., E.Elizalde and K.Kirsten: { Heat kernel 
coefficients of the Laplace operator on the D-dimensional ball}, 
\jmp{37}{96}{895}.}
\ref{BGKE}{Bordag,M., B.Geyer, K.Kirsten and E.Elizalde,: { Zeta function
determinant of the Laplace operator on the D-dimensional ball}, 
\cmp{179}{96}{215}.}
\ref{BKD}{Bordag,M., K.Kirsten,K. and Dowker,J.S.: Heat kernels and
functional determinants on the generalized cone \cmp{182}{96}{371}.}
\ref{Branson}{Branson,T.P.: Conformally covariant equations on differential
forms \cpde{7}{82}{393-431}.}
\ref{BandG2}{Branson,T.P. and Gilkey,P.B. {\it Comm. Partial Diff. Eqns.}
{\bf 15} (1990) 245.}
\ref{BGV}{Branson,T.P., P.B.Gilkey and D.V.Vassilevich {\it The Asymptotics
of the Laplacian on a manifold with boundary} II, hep-th/9504029.}
\ref{BCZ1}{Bytsenko,A.A, Cognola,G. and Zerbini, S. : Quantum fields in
hyperbolic space-times with finite spatial volume, hep-th/9605209.}
\ref{BCZ2}{Bytsenko,A.A, Cognola,G. and Zerbini, S. : Determinant of 
Laplacian on a non-compact 3-dimensional hyperbolic manifold with finite
volume, hep-th /9608089.}
\ref{CandH2}{Camporesi,R. and Higuchi, A.: Plancherel measure for $p$-forms
in real hyperbolic space, \jgp{15}{94}{57-94}.} 
\ref{CandH}{Camporesi,R. and A.Higuchi {\it On the eigenfunctions of the 
Dirac operator on spheres and real hyperbolic spaces}, gr-qc/9505009.}
\ref{CandW}{Candelas,P. and Weinberg,S. \np{257}{84}{397}}
\ref{CEJM}{Chamblin,A.,Emparan,R.,Johnson,C.V. and Myers,R.C.
 {\it Large N Phases. Gravitational Instantons and the Nuts and
Bolts of AdS Holography} \break hep-th/9808177.}
\ref{ChandD}{Chang, Peter and J.S.Dowker :Vacuum energy on orbifold factors
of spheres, \np{395}{93}{407}.}
\ref{cheeg1}{Cheeger, J.: Spectral Geometry of Singular Riemannian Spaces.
\jdg {18}{83}{575}.}
\ref{cheeg2}{Cheeger,J.: Hodge theory of complex cones {\it Ast\'erisque} 
{\bf 101-102}(1983) 118-134}
\ref{Chou}{Chou,A.W.: The Dirac operator on spaces with conical 
singularities and positive scalar curvature, \tams{289}{85}{1-40}.}
\ref{CandT}{Copeland,E. and Toms,D.J.: Quantized antisymmetric 
tensor fields and self-consistent dimensional reduction 
in higher-dimensional spacetimes, \break \np{255}{85}{201}}
\ref{DandH}{D'Eath,P.D. and J.J.Halliwell: Fermions in quantum cosmology 
\prD{35}{87}{1100}.}
\ref{cheeg3}{Cheeger,J.:Analytic torsion and the heat equation. \aom{109}
{79}{259-322}.}
\ref{DandE}{D'Eath,P.D. and G.V.M.Esposito: Local boundary conditions for 
Dirac operator and one-loop quantum cosmology \prD{43}{91}{3234}.}
\ref{DandE2}{D'Eath,P.D. and G.V.M.Esposito: Spectral boundary conditions 
in one-loop quantum cosmology \prD{44}{91}{1713}.}
\ref{CrandD}{Critchley,R. and Dowker,J.S. \jpa{14}{81}{1943}; 
{\it ibid} {\bf 15} (1982) 157.} 
\ref{Dow1}{Dowker,J.S.: Effective action on spherical domains, 
\cmp{162}{94}{633}.}
\ref{Dow8}{Dowker,J.S. {\it Robin conditions on the Euclidean ball} 
MUTP/95/7; hep-th\break/9506042. {\it Class. Quant.Grav.} to be published.}
\ref{Dow9}{Dowker,J.S. {\it Oddball determinants} MUTP/95/12; 
hep-th/9507096.}
\ref{Dow10}{Dowker,J.S. {\it Spin on the 4-ball}, 
hep-th/9508082, {\it Phys. Lett. B}, to be published.}
\ref{dow11}{Dowker,J.S. {\it Vacuum energy on the squashed 3-sphere}
in {\it Quantum Gravity} ed by S.C.Christensen 1984 IOP, Bristol.}
\ref{DandA2}{Dowker,J.S. and J.S.Apps, {\it Functional determinants on 
certain domains}. To appear in the Proceedings of the 6th Moscow Quantum 
Gravity Seminar held in Moscow, June 1995; hep-th/9506204.}
\ref{DABK}{Dowker,J.S., Apps,J.S., Bordag,M. and Kirsten,K.: Spectral 
invariants for the Dirac equation with various boundary conditions 
{\it Class. Quant.Grav.} to be published, hep-th/9511060.}
\ref{EandR}{E.Elizalde and A.Romeo : An integral involving the
generalized zeta function, {\it International J. of Math. and 
Phys.} {\bf13} (1994) 453.}
\ref{ELV2}{Elizalde, E., Lygren, M. and Vassilevich, D.V. : Zeta function 
for the laplace operator acting on forms in a ball with gauge boundary 
conditions. hep-th/9605026}
\ref{ELV1}{Elizalde, E., Lygren, M. and Vassilevich, D.V. : Antisymmetric
tensor fields on spheres: functional determinants and non-local
counterterms, \jmp{}{96}{} to appear. hep-th/ 9602113}
\ref{Kam2}{Esposito,G., A.Y.Kamenshchik, I.V.Mishakov and G.Pollifrone: 
Gravitons in one-loop quantum cosmology \prD{50}{94}{6329}; 
\prD{52}{95}{3457}.}
\ref{Erdelyi}{A.Erdelyi,W.Magnus,F.Oberhettinger and F.G.Tricomi {\it
Higher Transcendental Functions} Vol.I McGraw-Hill, New York, 1953.}
\ref{Esposito}{Esposito,G.: { Quantum Gravity, Quantum Cosmology and 
Lorentzian Geometries}, Lecture Notes in Physics, Monographs, Vol. m12, 
Springer-Verlag, Berlin 1994.}
\ref{Esposito2}{Esposito,G. {\it Nonlocal properties in Euclidean Quantum
Gravity}. To appear in Proceedings of 3rd. Workshop on Quantum Field Theory
under the Influence of External Conditions, Leipzig, September 1995; 
gr-qc/9508056.}
\ref{EKMP}{Esposito G, Kamenshchik Yu A, Mishakov I V and Pollifrone G.:
One-loop Amplitudes in Euclidean quantum gravity.
\prD{52}{96}{3457}.}
\ref{ETP}{Esposito,G., H.A.Morales-T\'ecotl and L.O.Pimentel {\it Essential
self-adjointness in one-loop quantum cosmology}, gr-qc/9510020.}
\ref{FORW}{Forgacs,P., L.O'Raifeartaigh and A.Wipf: Scattering theory, 
U(1) anomaly and index theorems for compact and non-compact manifolds 
\np{293}{87}{559}.}
\ref{GandM}{Gallot S. and Meyer,D. : Op\'erateur de coubure et Laplacian
des formes diff\'eren-\break tielles d'une vari\'et\'e riemannienne 
\jmpa{54}{1975}{289}.}
\ref{Gibb}{Gibbons, G.W. \aop {125}{80}{98}.}
\ref{Gilkey1}{Gilkey, P.B, Invariance theory, the heat equation and the
Atiyah-Singer index theorem, 2nd. Edn., CRC Press, Boca Raton 1995.}
\ref{Gilkey2}{Gilkey,P.B.:On the index of geometric operators for 
Riemannian manifolds with boundary \aim{102}{93}{129}.}
\ref{Gilkey3}{Gilkey,P.B.: The boundary integrand in the formula for the 
signature and Euler characteristic of a manifold with boundary 
\aim{15}{75}{334}.}
\ref{Grubb}{Grubb,G. {\it Comm. Partial Diff. Eqns.} {\bf 17} (1992) 
2031.}
\ref{GandS1}{Grubb,G. and R.T.Seeley \cras{317}{1993}{1124}; \invm{121}{95}
{481}.}
\ref{GandS}{G\"unther,P. and Schimming,R.:Curvature and spectrum of compact
Riemannian manifolds, \jdg{12}{77}{599-618}.}
\ref{Hitch}{Hitchin,N. \aim {14}{74}{1}.}
\ref{HFP}{Hu,B-.L, Fulling,S.A. and Parker, L. \prD{8}{73}{2377}.}
\ref{Hu}{Hu,B-.L. \prD{8}{73}{1048}.}
\ref{HHP}{Hawking,S.W, Hunter,C.J. and Page, D.N {\it Nut Charge,
Anti-de Sitter Space and Entropy} hep-th/9809035.}
\ref{IandT}{Ikeda,A. and Taniguchi,Y.:Spectra and eigenforms of the 
Laplacian
on $S^n$ and $P^n(C)$. \ojm{15}{1978}{515-546}.}
\ref{IandK}{Iwasaki,I. and Katase,K. :On the spectra of Laplace operator
on $\La^*(S^n)$ \pja{55}{79}{141}.}
\ref{JandK}{Jaroszewicz,T. and P.S.Kurzepa: Polyakov spin factors and 
Laplacians on homogeneous spaces \aop{213}{92}{135}.}
\ref{Kam}{Kamenshchik,Yu.A. and I.V.Mishakov: Fermions in one-loop quantum 
cosmology \prD{47}{93}{1380}.}
\ref{KandM}{Kamenshchik,Yu.A. and I.V.Mishakov: Zeta function technique for
quantum cosmology {\it Int. J. Mod. Phys.} {\bf A7} (1992) 3265.}
\ref{KandC}{Kirsten,K. and Cognola.G,: { Heat-kernel coefficients and 
functional determinants for higher spin fields on the ball} \cqg{13}{96}
{633-644}.}
\ref{Levitin}{Levitin,M.: { Dirichlet and Neumann invariants for Euclidean
balls}, {\it Diff. Geom. and its Appl.}, to be published.}
\ref{Lindel}{Lindel\"of.E. {\it Le Calcul des R\'esidus},
Gauthier-Villars, Paris, 1905.}
\ref{Luck}{Luckock,H.C.: Mixed boundary conditions in quantum field theory 
\jmp{32}{91}{1755}.}
\ref{MandL}{Luckock,H.C. and Moss,I.G,: The quantum geometry of random 
surfaces and spinning strings \cqg{6}{89}{1993}.}
\ref{Ma}{Ma,Z.Q.: Axial anomaly and index theorem for a two-dimensional 
disc 
with boundary \jpa{19}{86}{L317}.}
\ref{Mcav}{McAvity,D.M.: Heat-kernel asymptotics for mixed boundary 
conditions \cqg{9}{92}{1983}.}
\ref{MandV}{Marachevsky,V.N. and D.V.Vassilevich {\it Diffeomorphism
invariant eigenvalue \break problem for metric perturbations in a bounded 
region}, SPbU-IP-95, \break gr-qc/9509051.}
\ref{Milton}{Milton,K.A.: Zero point energy of confined fermions 
\prD{22}{80}{1444}.}
\ref{MandS}{Mishchenko,A.V. and Yu.A.Sitenko: Spectral boundary conditions 
and index theorem for two-dimensional manifolds with boundary 
\aop{218}{92}{199}.}
\ref{Moss}{Moss,I.G.: Boundary terms in the heat-kernel expansion 
\cqg{6}{89}{759}.}
\ref{MandP}{Moss,I.G. and S.J.Poletti: Conformal anomaly on an Einstein 
space 
with boundary \pl{B333}{94}{326}.}
\ref{MandP2}{Moss,I.G. and S.J.Poletti \np{341}{90}{155}.}
\ref{NandOC}{Nash, C. and O'Connor,D.J.: Determinants of Laplacians, the 
Ray-Singer torsion on lens spaces and the Riemann zeta function 
\jmp{36}{95}{1462}.}
\ref{NandS}{Niemi,A.J. and G.W.Semenoff: Index theorem on open infinite 
manifolds \np {269}{86}{131}.}
\ref{NandT}{Ninomiya,M. and C.I.Tan: Axial anomaly and index thorem for 
manifolds with boundary \np{245}{85}{199}.}
\ref{norlund2}{N\"orlund~N. E.:M\'emoire sur les polynomes de Bernoulli.
\am {4}{21} {1922}.}
\ref{Okada}{Okada,Y. \cqg{}{86}{221}.}
\ref{Poletti}{Poletti,S.J. \pl{B249}{90}{355}.}
\ref{RandT}{Russell,I.H. and Toms D.J.: Vacuum energy for massive forms 
in $R^m\times S^N$, \cqg{4}{86}{1357}.}
\ref{RandS}{R\"omer,H. and P.B.Schroer \pl{21}{77}{182}.}
\ref{SandS}{Shen, T.C. and Sobczyk,J. \prD{36}{87}{397}.}
\ref{SHC}{Shen,T.C., Hu,B.L. and O'Connor,D.J. \prD{31}{85}{2401}.}
\ref{SandV}{Shtykov,N. and Vassilevich, D.V. \jpa {28}{95}{L37}.}

\ref{Trautman}{Trautman,A.: Spinors and Dirac operators on hypersurfaces 
\jmp{33}{92}{4011}.}
\ref{Vass}{Vassilevich,D.V.{Vector fields on a disk with mixed 
boundary conditions} gr-qc /9404052.}
\ref{Voros}{Voros,A.:
Spectral functions, special functions and the Selberg zeta function.
\cmp{110}{87}439.}
\ref{Ray}{Ray,D.B.: Reidemeister torsion and the Laplacian on lens
spaces \aim{4}{70}{109}.}
\ref{McandO}{McAvity,D.M. and Osborn,H. Asymptotic expansion of the heat kernel
for generalised boundary conditions \cqg{8}{91}{1445}.}
\ref{AandE}{Avramidi,I. and Esposito,G. Heat kernel asymptotics with 
generalised boundary conditions, hep-th/9701018.}
\ref{MandS}{Moss,I.G. and Silva P.J., Invariant boundary conditions for
gauge theories gr-qc/9610023.}
\ref{barv}{Barvinsky,A.O.\pl{195B}{87}{344}.}
\ref{krantz}{Krantz,S.G. Partial Differential Equations and Complex
Analysis (CRC Press, Boca Raton, 1992).}
\ref{treves}{Treves,F. Introduction to Pseudodifferential and Fourier 
Integral Operators,\break Vol.1, (Plenum Press,New York,1980).}
\ref{EandS}{Egorov,Yu.V. and Shubin,M.A. Partial Differential Equations
(Springer-Verlag, Berlin,1991).}
\ref{AandS}{Abramowitz,M. and Stegun,I.A. Handbook of Mathematical 
Functions (Dover, New York, 1972).}
\ref{ACNY}{Abouelsaood,A., Callan,C.G., Nappi,C.R. and Yost,S.A.\np{280}
{87}{599}.}
\ref{BGKE}{Bordag,M., B.Geyer, K.Kirsten and E.Elizalde, { Zeta function
determinant of the Laplace operator on the D-dimensional ball}, 
\cmp{179}{96}{215}.}

\end{putreferences}
\bye